\documentclass[a4paper,11pt]{article}
\pdfoutput=1 

\usepackage{jheppub} 

\usepackage[T1]{fontenc} 
\usepackage{amsmath}
\usepackage{physics}
\usepackage{tikz}
\usepackage{multirow}
\usepackage[dvipsnames]{xcolor}
\usepackage{caption}
\usepackage{ulem}
\usepackage{tabularx} 
\newcommand{\ghc}{\mathcal{G}_{\mathrm{HC}}}
\newcommand{\fgc}{f_{\mathrm{GC}}}
\newcommand{\fa}{f_{\mathrm{a}}}
\newcommand{\fsp}{f_{\mathrm{Sp}}}
\newcommand{\Lsp}{\Lambda_{\mathrm{Sp}}}

\newcommand{\Ngc}{N}
\newcommand{\SUNgc}{SU \left(2N + 3\right)_{\text{GC}}}
\newcommand{\Sp}{Sp(2\Ngc)}
\newcommand{\Suc}{SU(3)_{\mathrm{c}}}
\newcommand{\lag}{\mathcal{L}}

\newcommand{\commented}[1]{}

\title{\boldmath Emergent axion and Higgs boson from strong dynamics}

\author[a]{F. Goertz}
\author[a,b]{and A. Incrocci}
\affiliation[a]{Max-Planck-Institut für Kernphysik, Saupfercheckweg 1, 69117 Heidelberg, Germany}
\affiliation[b]{Institute for Theoretical Physics, Karlsruhe Institute of Technology, 76131 Karlsruhe, Germany}

\emailAdd{florian.goertz@mpi-hd.mpg.de}
\emailAdd{andrea.incrocci@kit.edu}

\abstract{We propose a unified model that simultaneously addresses the hierarchy problem and the strong CP problem by considering the most simple
fundamental composite Higgs setup and showing that it can feature a viable emergent axion that could be accessible at collider experiments.
To maintain a low decay constant, and therefore address the hierarchy problem, while simultaneously avoiding experimental bounds on the axion couplings, we increase the axion mass via additional small instanton contributions coming from a new hidden gauge sector with a confinement scale larger than $\Lambda_{\text{QCD}}$.

Specifically, the axion will be identified with the CP-odd scalar singlet contained in the $SU(4)/Sp(4)$ coset of the minimal fundamental composite Higgs model. Both the SM color group and the additional hidden sector group are then embedded into a larger non-Abelian \textit{grandcolor} group, such that the topological angles of the two sectors are guaranteed to agree at tree-level. Beyond that, we show that radiative corrections and other CP-violating sources can be controlled. 
After examining the field content and the gauge structure of the model, we analyze the pNGB spectrum, potential, and couplings, as well as the resulting phenomenology. We focus in particular on the axion potential, to understand under which conditions the CP-odd scalar singlet can solve the strong CP problem while maintaining a naturally low compositeness scale, identifying interesting viable parameter space for an axion in the GeV range.}

\begin{document} 
\maketitle
\flushbottom

\section{Introduction}
\label{sec:introduction}

During the five decades since its formulation in the 1970s, the Standard Model (SM) of particle physics has provided an extremely successful description of nature. It has allowed us to describe three of the four fundamental forces, i.e.~the electromagnetic, the strong and the weak interaction, in a relatively simple framework and with an unprecedented level of accuracy. Yet, despite its success, there are strong reasons to conclude that the SM is not the final theory, but rather an effective description to be completed by a more fundamental model at shorter length scales. The shortcomings of the SM can be divided into two categories: \textit{problems}, that is experimentally well established phenomena that the SM can not accommodate, and \textit{puzzles}, i.e. aspects that while not inconsistent with the SM, lack a satisfactory explanation from first principles or appear fine-tuned and might find a more \textit{natural} origin in an even more fundamental theory. To name a few, among the \textit{problems} there are dark matter, baryogenesis, and neutrino masses,  while among the \textit{puzzles}, despite their name, there are the strong CP problem, the hierarchy problem, possible grand unification, and the flavor puzzle.

The great success of the SM can be traced back to two core ideas or guiding principles: \textit{unification} and \textit{symmetry}. This work focuses on two of the aforementioned puzzles, the strong CP problem and the hierarchy problem, and explores whether a \textit{unified} solution to both can emerge in terms of a larger and more fundamental \textit{symmetry}. We will therefore try to realize a model in which the axion \cite{Peccei:1977hh,Peccei:1977ur,Weinberg:1977ma,Wilczek:1977pj}, a leading candidate solution to the strong CP problem, emerges as a pseudo-Nambu-Goldstone boson (pNGB) from the {\it same} symmetry breaking that generates a composite Higgs boson \cite{Weinberg:1975gm, Kaplan:1983fs, Susskind:1978ms, Kaplan:1983sm,
Dugan:1984hq}, one of the most compelling solutions to the hierarchy problem (see also Refs. \cite{Panico:2015jxa, Contino:2010rs, Bellazzini:2014yua, Goertz:2018dyw} for more recent reviews).

The idea of unifying the Higgs and the axion has been explored earlier, however either without trying to address the hierarchy problem \cite{Redi:2012ad, Alanne:2018fns}, or without making use of the CP-odd singlet in the coset \cite{Gherghetta:2020ofz}. Since we are after a model in which the axion and the Higgs boson arise from the same symmetry breaking, we need to identify the axion scale, $f_a$, i.e. the scale at which the $U(1)_{\mathrm{PQ}}$ symmetry is spontaneously broken, with the Higgs compositeness scale, $f$. Therefore, the following relation has to hold
\begin{equation}
    f_a \equiv f = v \sin \theta,
\end{equation}
where $v$ is the electroweak (EW) scale, and $\theta$ is the misalignment angle of the vacuum of the composite sector w.r.t.~the EW one. Unfortunately, a standard QCD axion with a breaking scale around the EW scale \cite{Weinberg:1977ma, Wilczek:1977pj} is already ruled out by experimental data \cite{Hall:1981bc}. Current experimental bounds roughly require~\cite{Dolan:2022kul}
\begin{equation}
    f_a \gtrsim 10^{8} \; \text{GeV}.
\end{equation}

If this was the case, we could not provide a solution to the hierarchy problem, since significant tuning would be needed in the Higgs sector to maintain a large separation between $v$ and $f_a$. Evading experimental bounds while keeping $f_a$ sufficiently low is far from an easy task. To avoid constraints from processes like $K^+\rightarrow \pi^+ +a$ , we will either make the axion heavy enough to forbid this decay channel, or we will make it enough weakly coupled to avoid detection. This has already been explored extensively in the literature, see~\cite{Dine:1986bg,Flynn:1987rs,Holdom:1985vx,Holdom:1982ex,Tye:1981zy,Dimopoulos:1979pp,
Yang:1978gq} for early works, where the axion mass was increased thanks to additional contributions coming from small instantons. A particular strategy is to assume that the axion mass receives a second contribution from a hidden gauge sector with a confinement scale larger than $\Lambda_{\text{QCD}}$. In this way the axion mass can approximately be expressed as
\begin{equation}
    m_a^2f_a^2 \simeq \Lambda_{\text{QCD}}^4+\Lambda^{\prime 4}.
    \label{modified_axion_mass}
\end{equation}
For this to happen, the axion needs to couple to both sectors
\begin{equation}
    \mathcal{L}_{a} \supset \left(\bar{\theta}_c +\frac{a}{f_a} \right) \frac{g_c^2}{32\pi^2}G_{\mu\nu}^a \tilde{G}^{a\,\mu\nu} +  \left(\bar{\theta}_{c ^\prime} +\frac{a}{f_a} \right) \frac{{g_{c ^ \prime}}^2}{32\pi^2}G_{\mu\nu}^{\prime a} \tilde{G} ^{\prime a\,\mu\nu} ,
    \label{eq:lag_topological_term}
\end{equation}
and to solve the strong CP problem we have to identify a reason that ensures

\begin{equation}
    \bar{\theta}_c =\bar{\theta}_{c ^ \prime } ,
    \label{eq:theta_relation}
\end{equation}
up to corrections smaller than  $10^{-10}$. A different value of the two topological terms would imply that a single axion cannot eliminate both. In addition, since we assume $\Lambda ^ \prime>\Lambda_{\text{QCD}}$, the minimization of the axion potential would tend to cancel $\theta_{c ^\prime }$ and not $\theta_{c}$, thus not solving the strong CP problem. Many mechanisms have been proposed to explain both the origin of this hidden gauge sector and to justify the relation in \eqref{eq:theta_relation}. The best known solutions involve either a softly broken mirror symmetry of the SM (including its potential extension) \cite{Rubakov:1997vp,Berezhiani:2000gh,Gianfagna:2004je,Fukuda:2015ana,Hook:2014cda,Dimopoulos:2016lvn,Hook:2019qoh}, the embedding of $\Suc$ in an extended color group~\cite{Gherghetta:2016fhp,Valenti:2022tsc, Bedi:2024kxe, Kivel:2022emq}, or the product of multiple color groups $SU(3)^N$ \cite{Agrawal:2017ksf, Gaillard:2018xgk,Csaki:2019vte}~\footnote{See also \cite{Bedi:2025hbz,Gherghetta:2025fip, Co:2022aav} for other interesting ideas on heavy and composite axions.}.

Here, we follow the second approach and in particular the one outlined in \cite{Valenti:2022tsc}, i.e. we assume the existence of an enlarged non-Abelian color group, from now on referred to as \textit{grandcolor} (GC). In this way, the relation in \eqref{eq:theta_relation} is justified by the fact that the two $\theta$ parameters are unified at some scale $f_{\text{GC}}$ and we will ensure that no misalignment is introduced at lower energies. We then assume that this enlarged color group breaks into the SM one and a new confining group ${\cal G}_{c \prime}$, reproducing at low energy the correct SM gauge group.

Turning to the EW sector, in order to realize the Higgs as a pNGB, we will consider the simplest fundamental composite Higgs setup, based on the $SU(4)/Sp(4)$ coset~\cite{Koulovassilopoulos:1993pw, Katz:2005au, Lodone:2008yy, Gripaios:2009pe}, see also~\cite{Cacciapaglia:2020kgq,Cacciapaglia:2022zwt} for reviews.  For the UV completion of this sector we will, as usual, employ yet an additional confining gauge sector, hereafter referred to as \textit{hypercolor} (HC). We further assume that \textit{hypercolor} and \textit{grandcolor} are embedded in an even larger color group, called \textit{extended technicolor} (ETC) \cite{Eichten:1979ah,Dimopoulos:1979es,Lane:2002wv}. In this way, the fermions in the two sectors can couple via the exchange of ETC gauge bosons and all confining physics is originating from one simple gauge group. The key feature of this most simple 4D-completable coset is that, in addition to a $SU(2)_{\mathrm{L}} \cross SU(2)_{\mathrm{R}}$ scalar bidoublet, there is also a CP-odd singlet pNGB, which we will indicate as $\eta$.
Once we gauge the global $SU(2)_{\mathrm{L}}$ with the SM $SU(2)_{\mathrm{L}}$ and the third generator of the global $SU(2)_{\mathrm{R}}$ with hypercharge, the bidoublet can be identified with the Higgs, while under some circumstances the pseudoscalar could be considered as a regular QCD axion, as originally pointed out in \cite{Gripaios:2009pe}. However, in this original model, it was not possible to address both the hierarchy and strong CP problem at the same time, due to the incompatibility of a canonical invisible axion, requiring a large $f_a$, and a natural composite Higgs with a small compositeness scale $f$. Consequently no detailed study was performed. Relying on our modified framework, sketched above, this paper will be devoted to understanding under which conditions in this different setup the singlet can provide a solution to the strong CP problem and to what extent the strong CP problem and the hierarchy problem can be solved simultaneously.

The paper is organized as follows. In Section \ref{sec:model_setup} we introduce the field content and the gauge structure of the model. In Section \ref{sec:global_symmetries} we discuss the various global symmetries and their breaking, which allows us to identify all the pNGBs. Moving to Section \ref{sec:chiral_lagrangian}, we will present the chiral Lagrangian formalism, which will be used to study the potential of the pNGBs. Due to their importance for the hierarchy and strong CP problem, the discussion of the Higgs and the axion potentials is postponed to Section \ref{sec:axion_higgs_potential}. The couplings of the axion to the SM fields are investigated in Section \ref{sec:axion_couplings}, while Section \ref{sec:phenomenology} contains a brief analysis of the phenomenology of the model.

\section{Model setup}
\label{sec:model_setup}
Starting from the \textit{grandcolor} gauge dynamics of \cite{Valenti:2022tsc}, we extend it to include a strongly coupled (HC) sector for the composite Higgs. The full gauge group of the model is therefore assumed to be 
\begin{equation}
    \ghc \times \SUNgc \times SU(2)_{\text{L}} \times U(1)_{\text{Y}^\prime}.
\end{equation}
The SM gauge group is then recovered by a two-step breaking of $\SUNgc$ via the vacuum expectation values (vevs) of two new scalar fields. The first, $\Phi$, transforms in the adjoint representation of \textit{grandcolor}, while the second one, $\Xi$, transforms in the 2-index antisymmetric representation \cite{Li:1973mq,Valenti:2022tsc}. Assuming that both scalar fields acquire a vev of order $\fgc$, the breaking pattern can be synthetically written as
\begin{equation}
\begin{gathered}
 \label{eq:breaking_pattern}
    \mathcal{G}_{\text{HC}}  \times \SUNgc \times SU(2)_{\text{L}} \times U(1)_{\text{Y}^\prime}   \\
    \downarrow \expval{\Phi} \,, \expval{\Xi} \sim \fgc  \\
    \mathcal{G}_{\text{HC}} \times \Sp \times \underbrace{\Suc \times SU(2)_{\text{L}} \times U(1)_{\text{Y}}}_{{\cal G}_{\rm SM}}.
\end{gathered}
\end{equation}
The SM left and right-handed quarks transform in the fundamental and anti-fundamental representation of the \textit{grandcolor} group, respectively. It follows that, in order to fill the group representation, the SM fields must be accompanied by exotic grandcolor partners, hereafter referred to as $\psi$. In addition, in order to avoid gauge anomalies the hypercharge must be partially embedded in $\SUNgc$ and in $U(1)_{Y ^ \prime}$ \cite{Valenti:2022tsc}, while to avoid triviality $\Ngc$ must be even \cite{Witten:1982fp}. Although the hierarchy problem associated with the two new scalar fields $\Phi$ and $\Xi$ would be less severe than the SM one, given that $\fgc \gg v$, we nevertheless assume them to be composite operators arising from an even more fundamental dynamics. The scalar and SM-like field content of the model is presented in Table \ref{tab:SM_field_content}.

\begin{table}[t]
    \centering
    \renewcommand{\arraystretch}{1.4} 
    \begin{tabular}{|c|cccc|}
        \hline
        & $\ghc$ & $\SUNgc$ & $SU(2)_\text{L}$  & $U(1)_{\text{Y}^\prime }$  \\ \hline\hline
    $Q$   & $\mathbf{1}$   & $(\mathbf{2\Ngc+3})$        & $\mathbf{2}$  & $\frac{1}{4\Ngc+6}$\\
    $U$   & $\mathbf{1}$   & $\overline{(\mathbf{2\Ngc+3})}$  & $\mathbf{1}$  & $-\frac{1}{2}-\frac{1}{4\Ngc+6}$\\
    $D$   & $\mathbf{1}$   & $\overline{(\mathbf{2\Ngc+3})}$  & $\mathbf{1}$  & $+\frac{1}{2}-\frac{1}{4\Ngc+6}$\\
    $l$   & $\mathbf{1}$   & $\mathbf{1}$  & $\mathbf{2}$  & $-\frac{1}{2}$               \\
    $e$   & $\mathbf{1}$   & $\mathbf{1}$  & $\mathbf{1}$  & $1$                           \\ \hline
    $\Phi$               & $\mathbf{1}$  & $\mathbf{Adj}$ & $\mathbf{1}$   & 0           \\
    $\Xi$                & $\mathbf{1}$  & $\mathbf{A}_2$ & $\mathbf{1}$   & $\frac{1}{2\Ngc+3}$\\  \hline
    
    \end{tabular}
    \caption{SM-like and scalar field content of the model and charges under the gauge group. Below $\fgc$, the SM-like quarks $Q,\, U$, and $D$ will split into the fundamental representations of $\Suc$ and $\Sp$.}
    \label{tab:SM_field_content}
\end{table}

Moving to $\ghc$, a classification of all possible purely fermionic UV complete models has been carried out in \cite{Ferretti:2013kya}. There, the authors have identified a set of conditions required for a viable model, among which are the absence of gauge anomalies, custodial symmetry and the presence of top partners.  Keeping in mind that we want to work with the $SU(4)/Sp(4)$ composite Higgs model, in order to have top partners we need to have some fermions charged under both \textit{hypercolor} and \textit{grandcolor}. A related scenario to ours has been studied in \cite{Gherghetta:2020ofz}, however both the color group and the nature of the axion differ from the one of our model. Nevertheless, as in those scenarios, we assume the existence of fermions, denoted by $\omega$ and $\chi$,  transforming in two different representations of $\mathcal{G}_{\text{HC}}$, namely $\Psi_{\mathrm{HC}} \in N_{\omega} R_{\omega} + N_{\chi}R_{\chi}$, where $N_{\omega, \chi}$ indicates the number of degrees of freedom in each representation, $R_{\omega,\chi}$. Their charges under the full gauge group are given in Table \ref{tab:HC_field_content}. The $\omega$ fermions, being charged under the EW gauge group, will constitute the composite Higgs, while the $\chi$ fermions, carrying grandcolor charge, are needed for the top partners. Following the nomenclature of \cite{Belyaev:2016ftv}, we are left with two possibilities for $\ghc$, that is the models M8 and M9, which are reported in Table \ref{tab:ghc_structures}.
With this field content, it can be shown \cite{Ferretti:2013kya} that the global symmetry is
\begin{equation}
    G_F^{\mathrm{HC}}=SU(4)_{\omega} \times SU(4\Ngc+6)_{\chi} \times U(1),
    \label{global_symmetry_hc}
\end{equation}
where the $SU(4\Ngc+6)_{\chi}$ symmetry comes from the $4\Ngc+6$ Weyl-fermions $\chi$ charged under GC, and
the additional $U(1)$ symmetry comes from the ABJ-anomaly-free combination of the two $U(1)_{\omega/\chi}$, associated with axial rotations of $\omega$ and $\chi$, respectively. Below the confinement scale of the \textit{hypercolor} sector, this global symmetry will be spontaneously broken, thus producing numerous pNGBs, among which the Higgs and the axion. For the \textit{hypercolor} group to confine in the IR, the gauge dynamics and the field content has to be such as to guarantee asymptotic freedom. In addition, one might desire a \textit{hypercolor} dynamics that lies near, but outside, the conformal window, in order to realize a mass mechanism \textit{à la} Partial Compositeness (see \cite{Kaplan:1991dc, Contino:2003ve, Agashe:2004rs} for original ideas and \cite{ Barnard:2013zea, Ferretti:2013kya, Ferretti:2014qta, Cacciapaglia:2014uja, Sannino:2016sfx, Cacciapaglia:2017cdi, Cacciapaglia:2020kgq} for realizations in UV complete models) or \textit{à la} walking technicolor \cite{Holdom:1981rm, Yamawaki:1985zg}.
The study of the conformal window \cite{Ryttov:2007sr,Dietrich:2006cm,Sannino:2009za,Appelquist:2007hu,DelDebbio:2010zz,Erdmenger:2020flu,Goertz:2024dnz} of this model is out of the scope of this paper and here we focus only on asymptotic freedom.  

\begin{table}[t]
    \centering
     \renewcommand{\arraystretch}{1.4} 
    \begin{tabular}{|c|cc|c|}
    \hline
     $\mathcal{G_{\text{HC}}}$ & $\omega$ & $\chi$ & Model \\ \hline\hline
    $Sp(2N_{\text{HC}})$  & $4\times \mathbf{F}$ & $(4 \Ngc +6) \times \mathbf{A}_2$ & M8 \\
    $SO(N_{\text{HC}})$   & $4\times \mathbf{Spin}$ & $(4 \Ngc +6)\times \mathbf{F}$ & M9 \\ \hline
    \end{tabular}
    \caption{Possible gauge groups for $\mathcal{G}_{\text{HC}}$ and new fermions charged under them. $\mathbf{F,\,A_{2}}$, $\textbf{Spin}$ denote the fundamental, two-index antisymmetric and spinorial representation, respectively. See text for details.}
    \label{tab:ghc_structures}
\end{table}

A second requirement comes from the dynamics of $\Sp$. To increase the axion mass, we need to ensure that the $\Sp$ charged fermions, which before we introduced as $\psi$, will confine at a scale $\Lsp \equiv 4\pi \fsp /\sqrt{N}$ \cite{Bedi:2024kxe} higher than $\Lambda_{\text{QCD}}$. More generally, in order to avoid experimental bounds, we require $\Lsp$ to be higher than the EW scale. We also know that $\Sp$ and $\Suc$ unify at a scale $\fgc$, therefore at such scale their gauge coupling has to be equal. We recall the expressions for the one loop RG-evolution of a gauge coupling $\alpha$, for the 1-loop $\beta$-function, as well as for the confinement scale $\Lambda$, 
\begin{equation}
    \frac{4\pi}{\alpha(\mu^2)}-\frac{4\pi}{\alpha(Q^2)}=\beta_0\text{ln}\left(\frac{\mu^2}{Q^2} \right).
\end{equation}
\begin{equation}
    \beta^{0}=\frac{11}{3}C_2(\mathbf{Ad})-\frac{2}{3} \sum_{i }N_{R_i}T(R_i),
\end{equation}
\begin{equation}
    \Lambda^2 \equiv \mu^2 \exp\left[ -\frac{4\pi}{\beta_0 \alpha(\mu^2)}\right],
\end{equation}
where $T(R_i)$ is the Dynkin index of the representation $R_i$, $N_{R_i}$ the number of corresponding Weyl fermions, and $C_2(\mathrm{Ad})$ the quadratic Casimir operator. Their values for the different groups are reported in the Appendix in Tables \ref{tab:dynkin_casimir_1},~\ref{tab:dynkin_casimir_2},~and \ref{tab:dynkin_casimir_3}  \cite{Erdmenger:2020flu}.
The constraints mentioned above then lead to
\begin{gather}
    \beta_0^{\mathrm{HC}}  > 0, \label{eq:hc_free} \\
    \beta_0^{\mathrm{Sp}}  > 0, \\
    \alpha_{\mathrm{c}}\left(\fgc^2 \right) = \alpha_{\mathrm{Sp}}\left(\fgc^2\right), \\
    \Lsp\geq v,
\end{gather}
where $\beta_0^{\mathrm{HC}}, \beta_0^{\mathrm{Sp}}$ are the one loop $\beta$-functions of $\ghc$ and $\Sp$, respectively, $\alpha_{\mathrm{c}}\left(\fgc^2 \right)$ and $\alpha_{\mathrm{Sp}}\left(\fgc^2 \right)$ the fine structure constants of $\Suc$ and $\Sp$ evaluated at a scale $\mu^2=\fgc^2$, and finally $\Lsp$ is the confinement scale of $\Sp$. 

Assuming $\ghc=SO(11)$, we take the HC field content to be the one presented in Table~\ref{tab:HC_field_content}, such that with $4 \leq \Ngc \leq 6$ and $f_a  \in \left[2, 10^{3}\right]$ TeV all the conditions are satisfied for a natural decay constant. Additionally, for these parameters, the unification scale $\fgc$ is fixed to be within the range $\fgc \in \left[10^3, 10^{10}\right]$ TeV. This range is compatible with the bounds derived in Section \ref{sec:cp_violation}, where we discuss how possible sources of CP-violation, which might spoil our axion solution to the strong CP problem, can be avoided.
The field content shown in Table \ref{tab:HC_field_content} is not yet sufficient to generate composite partners for all three generations of SM quarks. The simplest approach to producing a flavor structure exclusively trough Partial Compositeness (PC) would be to introduce more than one (at least three) generations of $\chi$-fermions in the HC-sector as well. Unfortunately this turns out to be in conflict with the constraints discussed above. To generate the flavor structure in a minimal way, we will thus rely on the minimalistic realization of PC, where only the top quark is composite, while the masses of all the other fermions arise via bilinear interactions of SM fields and HC ones.

\begin{table}[t]
    \centering
    \renewcommand{\arraystretch}{1.4} 
    \begin{tabular}{|c|cccc|}
        \hline
        & $SO(N_{\text{HC}})$  & $\SUNgc$ & $SU(2)_\text{L}$  & $U(1)_{\text{Y}^{\prime}}$  \\ \hline\hline
    $\omega_{1,2}$ & $\mathbf{Spin}$ & $\mathbf{1}$        & $\mathbf{2}$  & 0 \\
    $\omega_{3}$   & $\mathbf{Spin}$ & $\mathbf{1}$        & $\mathbf{1}$  & $-\frac{1}{2}$ \\
    $\omega_{4}$   & $\mathbf{Spin}$ & $\mathbf{1}$        & $\mathbf{1}$  & $\frac{1}{2}$ \\ \hline
    $\chi_{a}$         & $\mathbf{F}$ & $\mathbf{F}$        & $\mathbf{1}$  &  $x$ \\
    $\chi_{\bar{a}}$   & $\mathbf{F}$ & $\bar{\mathbf{F}}$  & $\mathbf{1}$  &   $-x$ \\ \hline
    
    \end{tabular}
    \caption{Hypercolor content of the model and charges under the gauge groups.}
    \label{tab:HC_field_content}
\end{table}

\section{Global symmetries and pNGBs}
\label{sec:global_symmetries}

Due to the relatively large particle content, the model we are considering contains a considerable quantity of pNGBs arising from the spontaneous breaking of the various global symmetries in the theory. The purpose of this section is to identify all these symmetries and classify the pNGBs.

At scales higher than $f_{\text{GC}}$, the global symmetry associated with the dynamics of the HC-fermions is the one already reported in \eqref{global_symmetry_hc}, namely \cite{Ferretti:2013kya}
\begin{equation}
 \label{global_symmetry}
    G_F^{\text{HC}}= SU(4)_{\omega}\times SU(4\Ngc+6)_{\chi}\times U(1).
\end{equation}
At scales lower than $f_{\text{GC}}$, the dynamics of $Sp(2\Ngc)$ and $\Suc$ could lead to the breaking of the global $SU(4\Ngc+6)_{\chi}$, as could the vevs of the scalars $\Xi$ and $\Phi$. 
In order to simplify the discussion and given the corrections are not expected to be phenomenologically relevant at the TeV scale, we will ignore these effects.
It would be interesting to investigate the details of the bound states in the $\chi$ sector, which are expected to have masses of order $f_a$, below further condensation at $\Lsp$, which is however beyond the scope of this work.

Another global symmetry arises from the exotic partners of the SM quarks, the $\psi$'s, and it is associated with flavor rotations of the three generations of the four $\psi$ fermions before the confinement of $Sp(2\Ngc)$. In this case the symmetry is \cite{Valenti:2022tsc}
\begin{equation}
    G_F^{\psi} = SU(12)_{\psi}.
\end{equation}
The formation of the different fermion condensates, namely $\expval{\omega\omega}$, $\expval{\chi\chi}$, and $\expval{\psi\psi}$ at a scale $f_{\omega}, f_{\chi}$, and $ \fsp$, respectively, will spontaneously break $G_F^{\text{HC}}$ and $G_F^{\psi}$ down to
\begin{align}
    H_F^{\text{HC}} &= Sp(4)_{\omega} \times SO(4\Ngc+6)_{\chi}, \\
    H_F^{\psi} & = Sp(12).
\end{align}
To simplify the discussion, in the following we will assume
\begin{equation}
     f_{\mathrm{a}} = f_{\omega} = f_{\chi}.
\end{equation}
The $\expval{\omega\omega}$ condensate, breaking the $SU(4)_{\omega}$ flavor symmetry, transforms in the 2-index antisymmetric representation, i.e. in the $\mathbf{6}$ of $SU(4)$ \cite{Katz:2005au, Lodone:2008yy,Koulovassilopoulos:1993pw, Gripaios:2009pe}. The resulting breaking pattern is
\begin{equation}
    SU(4)\xrightarrow{}Sp(4), \quad \text{with } 5 \text{ pNGBs}.
\end{equation}
We will collectively indicate these pNGBs with $\pi_{\omega}$.  In addition, among the $\omega\omega$ mesons there will also be an additional gauge singlet, from now on indicated with $\sigma_{\omega}$. 

The $SU(4\Ngc+6)_{\chi}$ symmetry is broken by the $\expval{\chi\chi}$ condensate, which transforms in the 2-index symmetric representation, i.e. the $\mathbf{(2\Ngc+3)(4\Ngc+7)}$ of $SU(4\Ngc+6)$~(c.f.~\cite{Cacciapaglia:2015eqa}). In this case, the breaking pattern reads
\begin{equation}
    SU(4\Ngc+6) \xrightarrow{}SO(4\Ngc+6), \quad \text{with } 8\Ngc^2+26\Ngc+20 \text{ pNGBs}.
\end{equation}
Among the numerous pNGBs, some will be charged under $\Suc$, some under $\Sp$, and others under both or none. We will indicate them with $\pi_{\mathrm{c}}, \pi_\mathrm{Sp}$, $\pi_{\mathrm{c,Sp}}$, and $\pi_0^\chi$, respectively. 
As we will see in more detail in Section \ref{sec:chiral_lagrangian}, the ($Sp$ and color) charged pNGBs will acquire a mass due to gauge boson loops. Knowing that at scales smaller that $f_a$ we expect $\alpha_{\mathrm{Sp}} \gg \alpha_{c}$, we have $m_{\pi_{\mathrm{Sp}}} > m_{\pi_{\mathrm{c}}}$. It is interesting to notice that the dynamics of the lighter mesons, i.e. of $\pi_\mathrm{c}$ is associated to the breaking of a smaller symmetry contained in $SU(4\Ngc + 6)$, namely $SU(6)$, which is related to flavor rotations of the colored component of the $\chi$ alone. Another way to see the origin of this symmetry, is that below $\fsp$,  the confinement scale of $\Sp$, we can only have $Sp$-neutral mesons, precisely the $\pi_{\mathrm{c}}$. As in the $\omega$ coset, the $\chi\chi$ mesons feature yet another gauge singlet, which we will indicate with $\sigma_{\chi}$~\cite{Cacciapaglia:2015eqa}.

\begin{figure}[t]
    \centering
    \includegraphics[scale=0.6]{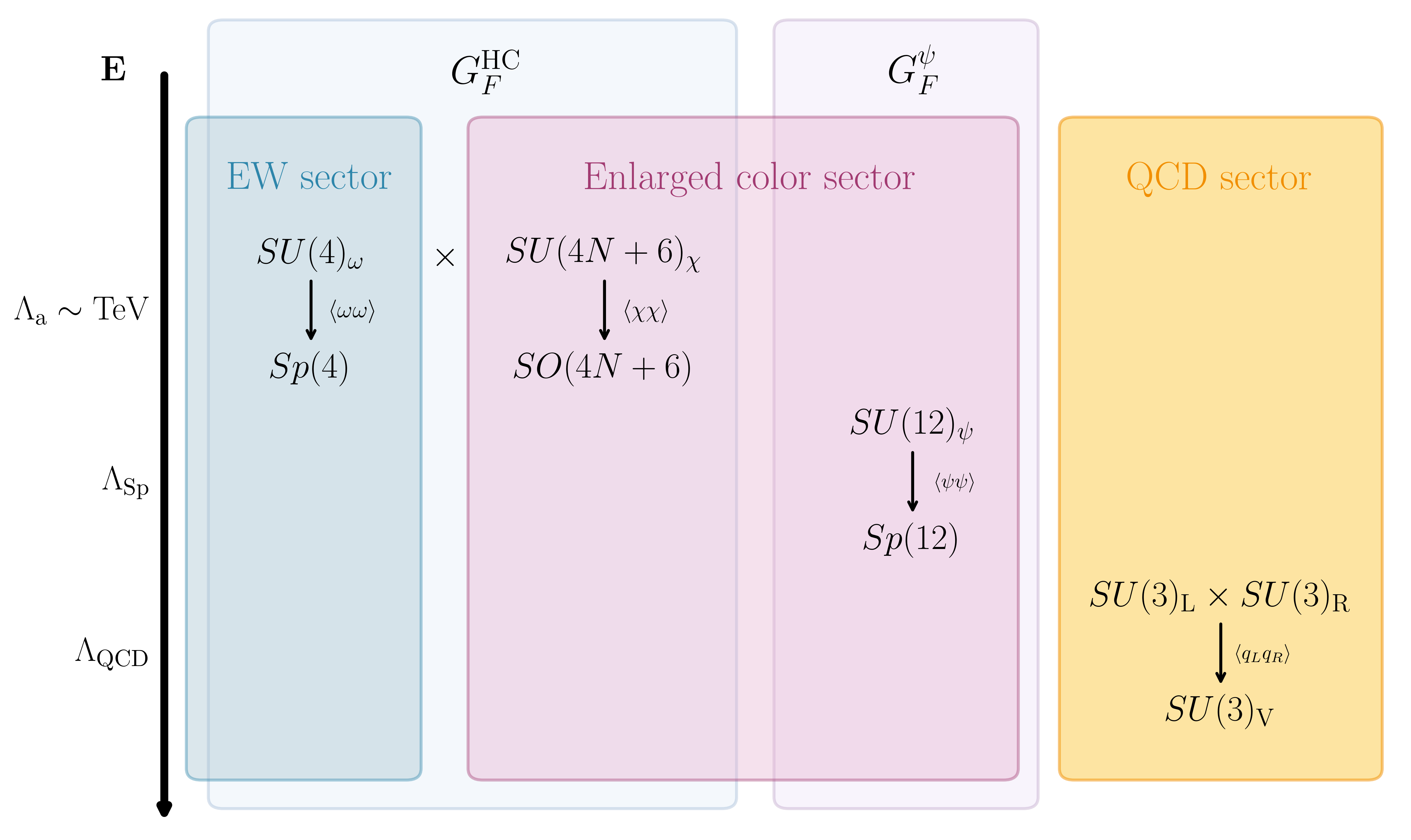}
    \caption{Pictorial representation of the various global symmetry breakings induced by the formation of the fermion condensates, leading to pNGBs.} 
    \label{fig:breaking_patterns}
\end{figure}

Lastly, the $SU(12)_{\psi}$ symmetry is broken by the $\expval{\psi\psi}$ condensate, that transforms in the 2-index antisymmetric representation, i.e in the $\mathbf{66}$ of $SU(12)$ \cite{Valenti:2022tsc} and the breaking patter is
\begin{equation}
    SU(12)\xrightarrow{}Sp(12), \quad \text{with } 65 \text{ pNGBs}.
\end{equation}
Again, we will indicate these pNGBs with $\pi_{\psi}$ and the additional singlet among the $\psi\psi$ mesons with $\sigma_{\psi}$. Among the $\pi_{\psi}$, we find a gauge singlet which we will indicate with $\eta_{B}$. This will in principle remain perfectly  massless, as it is associated with a $U(1)_{B}$ symmetry which remains unbroken. We will further comment on this pNGB in Section \ref{sec:axion_potential}. A schematic representation of the different global symmetries and their breaking is given in Figure \ref{fig:breaking_patterns}.

We have already commented how the $U(1)$ symmetry in $G_F^{\mathrm{HC}}$ is the ABJ anomaly free combination of the two axial symmetries $U(1)_{\omega}$ and $U(1)_{\chi}$ \cite{Belyaev:2016ftv}. Among the mesons described above, we therefore expect an additional pNGB we have so far not identified. This corresponds to a linear combination of the two mesonic states $\sigma_{\omega}$ and $\sigma_{\chi}$, while the orthogonal state acquires a mass due to large HC instanton effects. In turn, the pNGB also acquires a mass due to the current mass term of the $\chi$ fermions, breaking the $U(1)$ of the anomaly free combination.

The final step to study the pNGBs is to identify their charges under the gauge group. For this purpose we have to look at how the gauge group is embedded into the global symmetry. In the introduction we have seen that we would like to identify the third generator of the global $SU(2)_R \subset Sp(4)$ with $U(1)_{Y}$. 
However, to be able to assign the appropriate hypercharge to the top partners, we have to add a component along an additional global $U(1)_X$, as is common in composite Higgs constructions. Thus, we consider the embedding
\begin{equation}
\begin{gathered}
    Sp(4)\supset SU(2)_{\mathrm{L}} \cross SU(2)_{\mathrm{R}},  \\
    SO(4\Ngc+6)\supset SU(2\Ngc+3) \times U(1)_X\,, \\
    \label{gauge_embedding}
\end{gathered}
\end{equation} 
and, referring to the breaking pattern in \eqref{eq:breaking_pattern}, we then gauge the generator
\begin{equation}
    Y^{\prime} =T_R^3+X
\end{equation}
and identify
\begin{equation}
    Y =Y^{\prime} + Y_{\text{GC}}=T_R^3+X + Y_{\text{GC}},
\end{equation}
where $Y_{\text{GC}}$ is related to an additional $U(1)_{\text{GC}}$ originating from the breaking of the \textit{grandcolor} group, namely $\SUNgc \xrightarrow{\langle \Phi \rangle} \Suc \times SU(2N) \times U(1)_{\text{GC}}$.

\section{The chiral Lagrangian}
\label{sec:chiral_lagrangian}
We proceed with the introduction of the chiral Lagrangian formalism that will be used to describe the dynamics and potential of the pNGBs and the additional gauge singlets. The chiral formalism for similar classes of models has already been studied in the literature, see Refs. \cite{Alanne:2018fns, Belyaev:2016ftv, Cacciapaglia:2014uja, DeGrand:2016pgq, Sannino:2016sfx} for a detailed discussion. Our model contains three fermion condensates, where we can parameterize the pNGBs with the matrices $\Sigma_r$ and the additional gauge singlets (analogous to the QCD $\eta ^ \prime$) with $\Phi_r$, where $r = \omega, \, \chi, \,\psi $. As mentioned above, a linear combination of these singlets corresponds to the pNGB associated with the spontaneously broken extra anomaly-free $U(1)$ symmetry present in the model. Following the literature, we define 
\begin{equation}
    \Sigma_r= \exp \left\{ \frac{i2\sqrt{2}c_5}{f_r} \pi^a_r T^a_r   \right\}\cdot \Sigma_{0}^r, \qquad \Phi_r=\exp \left\{ \frac{ic_5 \sigma_r}{f_{\sigma_r}}  \right\},
    \label{eq:sigma_phi}
\end{equation}
where $T^a_r$ are the generators in the fundamental irrep normalized as $\Tr\left[ T^a_r T^b_r \right]=\delta^{ab}/2$, $f_r$ and $f_{\sigma_r}$ are the decay constants, $\Sigma_{0}^r$ is the gauge preserving vacuum and $c_5$ is a constant defined as 
\begin{equation}
    c_5=
\begin{cases}
  \sqrt{2} \text{ for real irrep,}\\
  1  \text{ for pseudoreal/complex irrep}.
\end{cases}
\label{eq:c5_constant}
\end{equation}
With these definitions, the lowest order chiral Lagrangian reads
\begin{equation}
    \lag_{\chi \mathrm{PT}} = \frac{f_r^2}{8c_5^2}\Tr \left[ \left(D_{\mu}\Sigma \right)^\dagger \left(D_{\mu}\Sigma \right) \right]+\frac{f_{\sigma_r}^2}{2c_5^2}\left( \partial_{\mu} \Phi_r \right)^{\dagger} \left( \partial^{\mu} \Phi_r \right).
    \label{kinetic_term}
\end{equation}

\subsection{The electroweak sector}
Starting with the pNGBs of the $SU(4)/Sp(4)$ coset, we parameterize $\Sigma_0^{\omega}$ in terms of the EW-preserving and EW-breaking vacua, respectively defined as 
\[
\begin{gathered}
\Sigma_B= \begin{pmatrix} 
i\sigma_2 & 0  \\
0 & -i\sigma_2 
\end{pmatrix},
\qquad 
\end{gathered}
\Sigma_H = \begin{pmatrix} 
0 & 1  \\
-1 & 0
\end{pmatrix}.
\]
The broken generators $X^a$ and the unbroken ones $T^a$ with respect to the EW preserving vacuum $\Sigma_B$ are 
\begin{align*}
X_{a} &= \frac{1}{2\sqrt{2}} \begin{pmatrix}
0 & \sigma_a \\
\sigma_a & 0
\end{pmatrix}, &
X_{4} &= \frac{i}{2\sqrt{2}} \begin{pmatrix}
0 & \mathbb{I} \\
-\mathbb{I} & 0
\end{pmatrix}, &
X_{5} &= \frac{1}{2\sqrt{2}} \begin{pmatrix}
\mathbb{I} & 0 \\
0 & -\mathbb{I}
\end{pmatrix},
\end{align*}
\begin{equation*}
T^{a}_{L} = \frac{1}{2} \begin{pmatrix}
\sigma_a & 0 \\
0 & 0
\end{pmatrix}, \quad
T^{a}_R = \frac{1}{2} \begin{pmatrix}
0 & 0 \\
0 & -\sigma_a^T
\end{pmatrix}, \quad
T^{a} = \frac{1}{2\sqrt{2}} \begin{pmatrix}
0 & i\sigma_a \\
-i\sigma_a & 0
\end{pmatrix}, \quad
T^{10} = \frac{1}{2\sqrt{2}} \begin{pmatrix}
0 & i\mathbb{I} \\
i\mathbb{I} & 0
\end{pmatrix},
\end{equation*}
where $a=1,2,3$ and $T^a_{L,R}$ correspond to the $SU(2)_{\mathrm{L}} \times SU(2)_{\mathrm{R}}$ subgroup. In terms of this unbroken subgroup, the pNGBs transforming in the $\mathbf{5}$ of $SU(4)$  decompose into $\mathbf{(2,2)\oplus(1,1)}$.
The EW symmetry will eventually be broken by the Higgs vev. For this reason, it is generally more convenient to take as the vacuum of the theory a linear combination of the two vacua defined above. In particular, we choose
\begin{equation}
    \Sigma_{0}^{\omega}=\cos\theta \Sigma_B + \sin\theta \Sigma_H=U_\theta\Sigma_B U_\theta^T, \qquad U_\theta=e^{i\sqrt{2}\theta X_2},
    \label{eq:omega_vacuum}
\end{equation}
where $\theta$ parametrizes the misalignment of the unbroken subgroup with respect to the EW embedding. The $SU(4)$ generator $X_2$ is then the one associated with the Higgs, while the singlet $\eta$ is associated with $X_5$. Using \eqref{eq:sigma_phi}, neglecting $\sigma_{\omega}$, and expanding in powers of $f_a$, the kinetic term \eqref{kinetic_term} becomes
\begin{equation}
    \lag_{\mathrm{kin}}^{\omega} \supset  \frac{1}{2}\left(\partial_{\mu} h\right)^2+\frac{1}{2}\left(\partial_{\mu}\eta\right)^2-\frac{1}{6f_a^2} \left( h\partial_{\mu}\eta +\eta \partial_{\mu}h \right)^2 + \frac{g_2^2}{4}f_a^2s_{\theta}^{2} \left( W_{\mu}^{+}W^{- \mu} +\frac{1}{2c_W} Z_{\mu}Z^{\mu} \right)    + \; ... \, ,
\end{equation}
where $s_{\theta}=\sin \theta$ and $c_W$ indicates the cosine of the Weinberg angle. From this expression we can immediately read off the gauge boson masses
\begin{equation}
    m_{W}^2=\frac{g_2^2}{4}f_a^2s_{\theta}^2,  \qquad m_Z^2=\frac{m_W^2}{c_W^2},
\end{equation}
so that we are lead to the identification $v=f_a s_{\theta}$. Since it will play a central role in the solution to the hierarchy problem, the discussion of the potential for the Higgs and $\eta$ fields is postponed to Section \ref{sec:axion_higgs_potential}.

\subsection{The (grand)color sector}
Moving to the $\expval{\chi\chi}$ condensate, we consider the vacuum
\begin{equation*}
    \Sigma_{0}^{\chi}=\begin{pmatrix} 
        0 & \mathbb{I} \\
        \mathbb{I}  & 0
        \end{pmatrix}\,,
\end{equation*}
where the identity is implicitly a $(2\Ngc+3)\times (2\Ngc+3)$ matrix. The generators of the unbroken subgroup $SU(2\Ngc+3)\times U(1)_X \subset SO(4\Ngc+6)$ are embedded as
\begin{equation*}
    S^a=\frac{1}{\sqrt{2}}\begin{pmatrix} 
        \lambda_a & 0\\
        0 & -\lambda_a^{T}
        \end{pmatrix},
\qquad
        X=x\begin{pmatrix} 
            \mathbb{I} & 0 \\
            0 & -\mathbb{I}
            \end{pmatrix},
\end{equation*}
where $\lambda_a$ are the generators of $\SUNgc$. In this coset, the first source of explicit symmetry breaking comes from the non-zero current mass of the fermions. Assuming for simplicity that the mass matrix is aligned with the $\SUNgc$\footnote{As mentioned before, here we are neglecting effects driven by the vevs of the two scalars and Yukawa interactions with the $\chi$ fermions, that are generally present.} preserving vacuum, we can write down the mass term
\begin{equation}
    \lag_m \supset m_{\chi} \; \chi^{T}. \begin{pmatrix} 
        0 & \mathbb{I}      \\
        \mathbb{I} & 0  
        \end{pmatrix} . \chi + \text{ h.c.} \,,
\end{equation}
where $m_{\chi}$ is real. Treating it as a spurion, we can build the invariant operator~\cite{Belyaev:2016ftv}
\begin{equation}
    V_m = \frac{f_a^2}{16}\Phi_{\chi}^2 \Tr[X^{\dagger}_{\chi}\Sigma_{\chi}] + \text{h.c.}  \, ,
\end{equation}
where $X_{\chi}=2B_{\chi} m_{\chi} \Sigma_{0}^{\chi}$ and $B_{\chi}$ is a dimensionful constant that can be computed on the lattice. In this case, the masses of all the $\chi$ mesons, both the pNGBs and the singlet, can approximately be expressed as
\begin{equation}
    m_{\pi_{\chi}}^2 \simeq 2 B_{\chi} m_{\chi}.
\end{equation}
On top of the current masses, a second contribution will come from gauge boson loops. We have anticipated in Section \ref{sec:global_symmetries} that among the numerous pNGBs, some will be charged under $\Sp$, some under $\Suc$, and some under both. We expect the gauge interaction to give a contribution \cite{Cacciapaglia:2015eqa} 
\begin{equation}
    \lag_{g} \sim C_gf_a^2 g^2 C_2 (\pi) \pi^2,
\end{equation}
where $C_g$ is an unknown $\mathcal{O}(1)$ parameter, $g$ the gauge coupling of either $\Sp$ or $\Suc$, and $C_2(\pi)$ the Casimir of the representation in which the pNGBs transforms. 

\subsection{The hidden color sector}

The pNGBs in the $\psi$ sector originating below the confinement scale of the $\Sp$ group, will play a crucial role in increasing the axion mass. Additionally, as we will see, they might lead to a problematic mixing with the Higgs boson. This coset requires therefore a careful study. Below $\fsp$, the chiral condensates $\expval{\psi_{q_u}\psi_{q_d}}=-\expval{\psi_{q_d}\psi_{q_u}}=\expval{\psi_{u}\psi_{d}}$ will form, inducing the breaking of the global $G_{\mathrm{F}}^{\psi}=SU(12)$ down to $Sp(12)$ \cite{Valenti:2022tsc}. Considering the twelve-dimensional basis-vector $\Psi =  (\psi_{q_u}^{(1)}, \psi_{q_d}^{(1)}, \ldots ,  \psi_u^{(1)}, \psi_d^{(1)}, \ldots  )$, with the superscript $i=1,2,3$ labeling generation space, we can write the vacuum as
\begin{equation}
    \Sigma_{0}^{\psi} = \mathrm{diag}\left(i\sigma_2, \,i\sigma_2, \, i\sigma_2,\,i\sigma_2,\, i\sigma_2,\,i\sigma_2\right).
    \label{eq:psi_vacuum}
\end{equation}
Following Ref. \cite{Bedi:2024kxe}, we estimate the condensate at confinement to be
\begin{equation}
    \expval{\psi\psi} \simeq \frac{\Ngc}{16\pi^2}\Lsp^3\Sigma_{0}^{\psi}\equiv \hat{\Lambda}_{\mathrm{Sp}}^3 \Sigma_0^{\psi},
\end{equation}
while we express the fluctuations around the condensate as
\begin{equation}
    \psi \psi \simeq  \frac{\Ngc}{16\pi^2}\Lsp^3 \left( \Sigma_{0}^{\psi}  + i\frac{2\sqrt{2}}{\fsp} \pi_{\psi} \cdot X_{\psi} \cdot \Sigma_0^{\psi} +  \; ...\right),
    \label{expansion_condensate}
\end{equation}
where following the traditional convention $X_{\psi}$ indicates the broken generators with respect to the vacuum in \eqref{eq:psi_vacuum}. In this sector, the sources of explicit symmetry breaking are given by the gauging of the EW group and the Yukawa interactions. In this particular case, 51 of the pNGBs are charged under $SU(2)_L \times U(1)_Y$ and, as we have seen before, will therefore acquire a positive mass squared of order $g_2^2\fsp^2$ \cite{Valenti:2022tsc}. The contribution due to the Yukawa interactions has already been investigated in \cite{Bedi:2024kxe}. Promoting the mass matrix originating from the Yukawa interaction to a $SU(12)_{\psi}$ spurion, one can include this breaking term in the chiral Lagrangian. We find that the leading correction to the mass of the charged pNGBs is proportional to $y_{\psi_u}^2 \fsp^2$, whereas for the neutral ones the correction is proportional to $y_{\psi_u} y_{\psi_d} \fsp^2$. Consequently, in the case of the pNGBs composed of 3rd-generation $\psi$ fermions, the top-Yukawa contribution dominates over the gauge one. As pointed out in \cite{Bedi:2024kxe}, this might be problematic because of the following reason. 

Among the charged pNGBs, some will carry the same charge as the Higgs boson $H$ (here the pNGB from the $\omega$ sector), which we will denote as $\pi_\psi^{\mathrm{H}}$. The Yukawa Lagrangian contains terms as 
\begin{equation}
   \lag_{\mathrm{Yuk}} \supset y_{\psi_t} \psi_{q_u}^{(3)} \psi_u^{(3)} H + \mathrm{h.c..}\,,
\end{equation}
which below the confinement scale generate via Eq.~\eqref{expansion_condensate} the mixing term  
\begin{equation}
{\cal L \supset }    y_{\psi_t} C_t \frac{\sqrt{N}}{4\pi} \Lsp^2 \pi^{\mathrm{H}}_{\psi} H + \mathrm{h.c.}.
\end{equation}
To obtain the mass eigenstate responsible for EWSB we have to diagonalize the mass-squared matrix 
\begin{equation}
    V \supset \left( H^{\dagger} \; \pi_{\psi}^{\mathrm{H}} \right)
    \begin{pmatrix}
        \mu_{H}^2 &  C_t y_{\psi_t} \frac{\sqrt{\Ngc}}{4\pi}\Lsp^2 \\
         C_t y_{\psi_t} \frac{\sqrt{\Ngc}}{4\pi}\Lsp^2 & \frac{\tilde{C}_t y_{\psi_t}^2 }{16\pi^2} \Lsp^2 
    \end{pmatrix}
    \begin{pmatrix}
        H \\
        \pi_{\psi}^{\mathrm{H}}
    \end{pmatrix},
\end{equation}
where $C_t$ and $\tilde{C}_t$ are unknown $\mathcal{O}(1)$ parameters, and where we have focused on the pNGB composed of third-generation $\psi$ fermions, since it is the one with the largest mixing with $H$. In order to have the electroweak scale below $\Lsp$, the mass matrix has to possess a small eigenvalue, or equivalently, its determinant must be small. If the leading contribution to the mass of the Higgs-like pion would come from the Yukawa interaction, then the (2,2) entry above would be large, similarly to the off-diagonal entries while it turns out that the (1,1) entry would also need to be large and tuned in such a way to generate a small eigenvalue in the mass-squared matrix. In this case the physical Higgs would predominantly have a component along $\pi_\psi^{\mathrm{H}}$ and consequently its projection along the $H$ bidoublet from the $\omega$ sector, which couples directly to the fermions, would be suppressed. If this is the case, then the top Yukawa coupling would be suppressed by the mixing angle, making it difficult to achieve a $\mathcal{O}(1)$ value. 

To avoid this, we will assume that the Yukawas of the third generation of the $\psi$ fermions, i.e. $y_{\psi_{t,b}}$, are smaller than their SM counterparts. 
Consequently, the off-diagonal terms in the mass matrix will be small and the $\pi_{\psi}^{\mathrm{H}}$ mass will be predominantly generated by EW gauge interactions, namely $m^2_{\pi_\psi^{\mathrm{H}}} \sim \tilde C_g g_{2}^2 \Lsp^2/16\pi^2$. For a sufficiently small exotic-top Yukawa interaction, the (1,1) entry ends up being smaller than the (2,2) one. In this case, we can integrate out the heavy Higgs-like pion field and the physical Higgs will mostly align with the $SU(4)$ bidoublet.
Assuming $C_t \simeq C_g \simeq 1$, the mixing angle reads
\begin{equation}
    \theta_{mix} \sim \frac{4\pi y_{\psi_t} \sqrt{N}}{g_2^2}.
\end{equation}
For consistency, asking $\theta_{mix} \ll 1$ implies $y_{\psi_t} \ll g_{2}^2/(4\pi\sqrt{\Ngc})\sim0.01$, as expected. 
In our model, such a suppression can be achieved by assuming that the mass of the top partner of $\psi_t$ is larger than that of the partner of the SM top. Indeed, in PC the size of the IR Yukawa interaction is suppressed by one power of the top partner mass $m_T$, namely (see, e.g., \cite{Panico:2011pw, Matsedonskyi:2012ym, Carmona:2014iwa})
\begin{equation}
    y \sim \frac{\lambda_L \lambda_R f_a}{m_T}\,,
\end{equation}
with $\lambda_{L,R}$ the linear-mixing coefficients, and therefore a top partner for $\psi_t$ approximately two orders of magnitude heavier than the one for the top would produce the sufficient suppression. 

\section{Axion coupling to the SM }
\label{sec:axion_couplings}
As in standard composite Higgs incarnations, the interaction of the \textit{hypercolor} sector with the SM-like one is a source of explicit symmetry breaking and therefore will induce couplings between the pNGBs of the hypercolor sector, among which is our axion $\eta$, and the SM-like fermions. To investigate how these couplings are generated, we need to specify how PC is implemented in this model (see, e.g., \cite{Barnard:2013zea, Ferretti:2013kya, Ferretti:2014qta, Cacciapaglia:2014uja, Sannino:2016sfx, Cacciapaglia:2017cdi, Cacciapaglia:2020kgq, Vecchi:2015fma, Goertz:2023nii}). While a detailed analysis, using the spurion method, is provided in \cite{Alanne:2018wtp}, here we report only the key points relevant for us.
As discussed in Section \ref{sec:model_setup}, due to the restrictions on the field content, we limit ourselves to top-PC and we assume that the mass of all the other quarks is generated by couplings that are bilinear in the SM-like fields. The top mass is then generated by the Lagrangian

\begin{equation}
    \lag_{t}^{\mathrm{PC}} = \lambda_{t_L}\bar{t}_L T_{t_L} +  \lambda_{t_R}\bar{t}_R T_{t_R}, 
\end{equation}
where $T_{t_L/ t_R}$ are fermionic composite operators containing two $\omega$ and one $\chi$ fermions, which carry the correct SM quantum numbers to couple to the respective SM-like fields. To include this symmetry breaking source in the chiral Lagrangian, we need to study how the SM fields can be embedded into $SU(4)$ representations. Here we will focus on the antisymmetric (A), i.e. the $\mathbf{6}$, and the symmetric (S), i.e. the $\mathbf{10}$.

Starting with the $\mathbf{6}$, under the unbroken $SU(2)_{\mathrm{L}}\times SU(2)_{\mathrm{R}}$ subgroup this representation decomposes into $(\textbf{2}, \Bar{\textbf{2}}) \oplus (\textbf{1},\textbf{1}) \oplus (\textbf{1},\textbf{1})$. We clearly have only one choice to embed $Q$, the bi-doublet, while for $t_R$ we have two. The projectors that select the components of the composite operators with the correct quantum numbers to couple to the respective standard model fields read
\begin{align}
P_1^Q &= \frac{1}{\sqrt{2}} \begin{pmatrix}
0 & 0 & 1 & 0 \\
0 & 0 & 0 & 0 \\
-1 & 0 & 0 & 0 \\
0 & 0 & 0 & 0 
\end{pmatrix},
&P_2^Q  = \frac{1}{\sqrt{2}}  \begin{pmatrix}
0 & 0 & 0 & 0 \\
0 & 0 & 1 & 0 \\
0 & -1 & 0 & 0 \\
0 & 0 & 0 & 0 
\end{pmatrix},
\nonumber \\
P_{t}^1  &= \frac{1}{\sqrt{2}}  \begin{pmatrix}
0 & 0 & 0 & 0 \\
0 & 0 & 0 & 0 \\
0 & 0 & 0 & 1 \\
0 & 0 & -1 & 0 
\end{pmatrix},
&P_{t}^2 = \frac{1}{\sqrt{2}}  \begin{pmatrix}
0 & 1 & 0 & 0 \\
-1 & 0 & 0 & 0 \\
0 & 0 & 0 & 0 \\
0 & 0 & 0 & 0 
\end{pmatrix},
\end{align}
where the index $i=1,2$ in $P^Q_i$ indicates the $SU(2)_L$ index. In general, it is convenient to take as the projector for the singlet a combination of the two above, in particular
\begin{equation}
    P_t =  A P_{t}^1 + BP_{t}^2, \quad \abs{A}^2+\abs{B}^2 =1.
    \label{top_right_spurion}
\end{equation}
Nevertheless, in what follows we choose $A=1, \, B=0$. As we will show later, such a choice is necessary if we want to avoid a contribution to the axion potential coming from the electroweak sector.

Moving to the $\mathbf{10}$, this representation decomposes into $(\textbf{2}, \Bar{\textbf{2}}) \oplus (\textbf{3},\textbf{1}) \oplus (\textbf{1},\textbf{3})$. We clearly have only one choice to embed $Q$, the bi-doublet, and also one choice for the singlet, the $(\mathbf{1}, \mathbf{3})$. In this case the embedding reads
\begin{equation}
P_1^Q = \frac{1}{\sqrt{2}} \begin{pmatrix}
0 & 0 & 1 & 0 \\
0 & 0 & 0 & 0 \\
1 & 0 & 0 & 0 \\
0 & 0 & 0 & 0 
\end{pmatrix},
\qquad
P_2^Q = \frac{1}{\sqrt{2}}  \begin{pmatrix}
0 & 0 & 0 & 0 \\
0 & 0 & 1 & 0 \\
0 & 1 & 0 & 0 \\
0 & 0 & 0 & 0 
\end{pmatrix},
\qquad
P_{t} = \frac{1}{\sqrt{2}}  \begin{pmatrix}
0 & 0 & 0 & 0 \\
0 & 0 & 0 & 0 \\
0 & 0 & 0 & 1 \\
0 & 0 & 1 & 0 
\end{pmatrix}.
\end{equation}
For these two representations, the LO operators for the top mass are given by \cite{Alanne:2018wtp} 
\begin{equation}
    \frac{\lambda_{t_L}\lambda_{t_R} f_a}{4\pi} \left( \bar{Q}^{\alpha}U \right) \times \begin{cases}
    C_{yA,1}\Tr\left[ P_Q^{\alpha}\Sigma^{\dagger} P_t^2 \Sigma^{\dagger} \right] +C_{yA,2}Tr\left[ P_Q^{\alpha}\Sigma^{\dagger}\right] \Tr\left[ P_t^2 \Sigma^{\dagger} \right], & \text{A}\\
    C_{yS}\Tr\left[ P_Q^{\alpha}\Sigma^{\dagger} P_t \Sigma^{\dagger} \right] & \, \text{S}\, .
  \end{cases}
  \label{eq:LO_top_mass}
\end{equation}
Expanding these expressions in powers of the fields we find the top mass
\begin{align}
 m_{t}^{A} = & \frac{\lambda_{t_L}\lambda_{t_R}}{4\pi}\left(C_{yA,1} +2C_{yA,2}\right) f_a c_{\theta}s_{\theta},  \\
 m_{t}^{S} = & \frac{\lambda_{t_L}\lambda_{t_R}}{4\pi} C_{yS} f_a c_{\theta}s_{\theta}\,,
\end{align}
and the expression of the top couplings with the Higgs and the axion, valid for both representations, 
\begin{equation}
    \label{singlet_top_coupling}
    \mathcal{L}_t \supset - m_t \left[\bar{t}t \left(  1 + \frac{c_{2\theta}}{c_{\theta}}\frac{h}{v} \right) -\frac{i}{f_a c_{\theta}}\bar{t}\gamma^5t \, \eta \right].
\end{equation}
We also find that no couplings between the axion and the other fermions are generated at LO for bilinear operators and neither are they at NLO, as this would require operators involving mass spurions, which are here absent \cite{Alanne:2018wtp}. We note that similar couplings are induced for the $\psi$ fermions. Following the normalization of \cite{Blasi:2023hvb}, we introduce the dimensionless $\eta$-top coupling strength $c_t = 1/c_{\theta}$. For all practical purposes, in our model we can assume $c_{\theta}\sim 1$, since $f_a \gg v$.

We can now study the couplings of the axion candidate $\eta$ with the gauge bosons, which are generated via the Wess-Zumino-Witten (WZW) anomaly. Following the normalization adopted in \cite{Belyaev:2016ftv}, they can be written as 
\begin{equation}
    \lag_{\mathrm{WZW}}\supset \frac{\sqrt{\alpha_{A^b}\alpha_{A^c}}}{4\sqrt{2}\pi}c_5 \frac{C_{A^bA^c}}{f_a}c^{bc}\eta \, \epsilon^{\mu\nu\alpha\beta}A^b_{\mu\nu}A^c_{\alpha\beta}\, ,
\end{equation}
where
\begin{equation}
    C_{A^b A^c}c^{bc} = \frac{d_{\omega}}{2} \Tr \left[ T^{\eta} \left\{S^b, S^c \right\} \right].
\end{equation}
In the expression above $d_{\omega}$ is the dimension of the representation of the $\omega$ fermions under HC, $c_5=1$ is the representation-dependent normalization constant defined in Eq.\eqref{eq:c5_constant}, $T^{\eta}$ is the generator associated with the singlet, and $S^{b,c}$ are the gauged generators. Interestingly the $\eta$ does not have a tree-level coupling with gluons, since the $\omega$ are not charged under color. In addition, at tree-level it will not couple to photons either, because $U(1)_{em}$ is fully embedded in $SU(4)$ \cite{Galloway:2010bp,Arbey:2015exa, Gripaios:2016mmi,Cacciapaglia:2021agf}. Nevertheless, an effective coupling to gluons, photons and the lighter fermions is generated at loop level via the $\eta$-top coupling. The interaction to gluons and photons takes the form ~\cite{Blasi:2023hvb} 
\begin{equation}
    \mathcal{L}_{a} \supset c_{\gamma \gamma} \frac{\alpha}{4\pi} \frac{a}{f_a}F\tilde{F} + c_{GG}\frac{\alpha_s}{4\pi}\frac{a}{f_a} G\tilde{G},
\end{equation}
where the effective couplings can be expressed as
\begin{subequations}
    \begin{align}
    c_{GG} & = \frac{1}{2}\sum_{f=\mathrm{quarks}}B_1\left(\frac{4m_f^2}{m_{\eta}^2}\right) c_f ,\\
    c_{\gamma\gamma}  & = \sum_f B_1\left(\frac{4m_f^2}{m_{\eta}^2}\right) N_c Q_f^2 c_f,
    \end{align}
    \label{eq:axion_photon_coupling}
\end{subequations}
and where 
\begin{equation}
   B_1\left(\frac{4m_f^2}{m_{\eta}^2}\right) \simeq 
   \begin{cases}
       1 & \text{if }  m_f \ll m_{\eta}\\
       - \frac{1}{12}\frac{m_{\eta}^2}{m_{f}^2} & \text{if }  m_{\eta} \ll m_f.
   \end{cases}
\end{equation}

In turns out \cite{Blasi:2023hvb}, that the two-loop contribution from light quarks can be more important than the one-loop contribution from the top. In particular, for $m_{\eta}\sim 10$\,GeV the main contributions come from charm and bottom loops. Even though axion-photon couplings are expected to be rather small in the model in general, the bounds on them could be relevant for $m_{\eta}<0.5\,$GeV -- see \cite{Caputo:2022mah, Fiorillo:2025yzf, Jaeckel:2017tud, Muller:2023vjm, Lucente:2020whw, Hoof:2022xbe} for supernovae bounds and  \cite{Dolan:2017osp, CHARM:1985anb, Riordan:1987aw, Blumlein:1990ay, NA64:2020qwq} for beam--dump bounds on the axion--photon couplings. A detailed study of the size of the axion--photon coupling is beyond the scope of this work, and we will therefore conservatively focus on the region $m_{\eta}>0.5\,$GeV in the phenomenology section. 
\section{The axion-Higgs potential}
\label{sec:axion_higgs_potential}
\subsection{Higgs potential}
The potential of the pNGBs of the $SU(4)/Sp(4)$ coset has already been extensively studied in the literature, see for example Refs. \cite{Katz:2005au, Galloway:2010bp,Alanne:2018wtp,Angelescu:2021pcd}. In this case the two sources of explicit symmetry breaking are given by the gauge and Yukawa interaction. In Ref.~\cite{Alanne:2018wtp} it was shown that, up to NLO, the potential takes the form
\begin{equation}
    V\left( \theta, h, \eta \right) = \sum_{i=1}^4 c_i f_i (\theta, h, \eta),
\end{equation}
where $f_i$ are specific functions of the misalignment angle and the fields. This expression holds only if the top spurions are embedded in a single $Sp(4)$ representation and when no mass spurion for $\omega$ is included. This justifies the choice $A=1,\, B=0$ made in \eqref{top_right_spurion}. In the case of massless $\omega$ fermions, we further have that $c_3$ and $c_4$ are zero. Focusing on the two relevant functions, they read
\begin{align}
    f_1  & = s_{\theta}^2 + 2c_{\theta}s_{\theta}\frac{h}{f} + c_{2\theta}\frac{h^2}{f^2}-s_{\theta}^2\frac{\eta^2}{f^2}-\frac{4}{3}c_{\theta}s_{\theta}\frac{h}{f}\frac{(h^2+\eta^2)}{f^2} + \ldots \, , \\
    f_2  & = s_{\theta}^4 +4c_{\theta}s_{\theta}^3 \frac{h}{f}+2s_{\theta}^2(1+2c_{2\theta})\frac{h^2}{f^2}-2s_{\theta}^4\frac{\eta^2}{f^2}-\frac{4}{3}c_{\theta}s_{\theta}(1-4c_{2\theta})\frac{h^3}{f^3} -\frac{20}{3}c_{\theta}s_{\theta}^3\frac{h \eta^2}{f^3} + \ldots \, .
\end{align}
Minimizing the potential gives us
\begin{equation}
    \frac{\partial V}{\partial \theta} = 0 \Rightarrow s_{\theta}^2=-\frac{c_1}{2c_2},
\end{equation}
so that, at the minimum, we can trade the two coefficients $c_1$ and $c_2$ for the misalignment angle $\theta$ and the Higgs mass $m_{h}$, namely
\begin{equation}
    c_1 = -\frac{f^2 m_h^2}{4c_{\theta}^2}.
\end{equation}
 By performing this substitution in the potential, one can check that $\eta$ remains massless. At the same time, we can also read the value of the trilinear coupling of the Higgs and the coupling between Higgs and $\eta$
\begin{align}
    g_{h^3} & = \frac{m_h^2}{2v} \frac{c_{2\theta}}{c_{\theta}}, \\
    g_{h\eta^2} & = - \frac{m_h^2}{2v}\frac{s_{\theta}^2}{c_{\theta}}.
\end{align}

\subsection{Axion potential}
\label{sec:axion_potential}
In our model, the solution to the strong CP problem relies on the fact that we can use the shift symmetry of $\eta$ to absorb the topological terms in \eqref{eq:lag_topological_term}. For this to be possible, the only source of explicit symmetry breaking has to come via QCD and Sp instantons. This would not be true if $\eta$ developed a vev due to some of the spurions we have considered so far. To include this possible contribution, in \eqref{eq:omega_vacuum} one should consider also a misalignment of the $SU(4)/Sp(4)$ vacuum along the direction of the $\eta$. This situation was already studied in \cite{Alanne:2018wtp} for this coset, where it was found that no spontaneous CP-violation via the vacuum misalignment is possible and that the only way to violate CP is to add CP-violating phases in the theory, which in our case are absent, due to vanishing current masses for $\omega$. We proceed by first discussing the full axion potential in the presence of the grandcolor extension and then how the possible CP violating sources are controlled.

To study the remaining possible sources for a potential of $\eta$, which could come from the new $\psi$ sector, i.e. the mixing with $\pi_{\psi}^{0}$, the neutral pions originating from the formation of the  $\expval{\psi\psi}$ condensate,  we closely follow \cite{Valenti:2022tsc, Bedi:2024kxe}. As pointed out by the authors of \cite{Valenti:2022tsc}, the dynamics of the neutral pions can be described in terms of a smaller flavor symmetry, subgroup of $SU(12)_{\psi}$, that is not explicitly broken by the EW interaction, namely
\begin{equation}
    SU(3)_q \cross SU(3)_u \cross SU(3)_d \cross U(1)_B.
\end{equation} 
This global symmetry is associated to independent flavor rotations of the $\psi$-fermions and it is broken by the formation of the chiral condensate down to
\begin{equation}
    SO(3)_q \cross SU(3)_V.
\end{equation}
Consequently, the neutral pNGBs can be described in terms of the matrices $\Sigma_L \in SU(3)_q/SO(3)_q$, $\Sigma_R \in SU(3)_u\cross SU(3)_d /SU(3)_V$ and $\eta_B \in U(1)_B$. Below the confinement of $Sp(2N_{\text{GC}})$ the Yukawa interactions will induce the potential \cite{Valenti:2022tsc,Bedi:2024kxe} 
\begin{equation}
    V \simeq  -\frac{N\Lsp^4}{(16\pi^2)^2}\Tr\left[ Y_\psi^u \Sigma_{\text{R}} (Y_\psi^d)^{T} \Sigma_{\text{L}}  \right] + \text{h.c.}\, \,
    \label{eq:axion_potential}
\end{equation}
where the axion enters in the expression via the Yukawa matrices, which in this case are those of the $\psi$ sector, not the SM ones. The numerical minimization of this potential was carried out in \cite{Valenti:2022tsc}, whose construction features the same LO axion-fermion couplings as our model (see \eqref{singlet_top_coupling}), and which found $\expval{\bar{\eta}}=0$. The authors also verified that the mixing between the axion and the neutral pNGBs induced by the potential in \eqref{eq:axion_potential} does not notably modify the axion couplings, even in the limit of $\fsp$ approaching $\fa$.

Moreover, flavor violating interactions arise at the HC scale, i.e.~$f_a$, 
and are generated by the operators
\begin{equation}
\frac{c_{ijkl}}{f_{a}^2}Q_{i}Q_{j}U_k D_l,
\label{eq:flavor_violation}
\end{equation}
where $i,j,k,l$ are flavor indices. As shown in \cite{Valenti:2022tsc}, this type of interactions can generate a correction to the axion potential \eqref{eq:axion_potential} of order
\begin{equation}
\delta V_0 \sim 16\pi^2 c \fsp^6 / (\Ngc^2f_a^2)\,.
\end{equation}
This term generally leads to misalignment of the axion vev away from zero of order $\expval{\bar{a}}/f_a \sim \Im\left[\delta V_0\right]/V_0$ \cite{Valenti:2022tsc}. Assuming order one coefficients, to ensure that the two topological angles remain smaller that $10^{-10}$ we would need to have
 \begin{equation}
    \fsp \lesssim 10^{-7} f_a.
    \label{eq:flavor_bound}
 \end{equation}
Flavor-violating interactions are a common problem in Composite Higgs models and are already heavily constrained by K and B meson physics~ \cite{UTfit:2007eik,Isidori:2013ez}. To avoid such operators, we thus assume that the composite operators posses some flavor symmetry and that therefore the bound in \eqref{eq:flavor_bound} can be relaxed.

 In conclusion, in our model the vev of $\eta$ precisely cancels $\theta_c$, effectively solving the strong CP problem. Furthermore, we can verify that the $\eta$ mass follows the relation (see also \cite{Valenti:2022tsc})
\begin{equation}
    m_{\eta}^2 \simeq \frac{2}{\Ngc} y_{\psi_u} y_{\psi_d} \frac{\fsp^4}{f_a^2c_{\theta}^2}.
    \label{eq:eta_mass_expression}
\end{equation}
Using as a benchmark the value of the SM Yukawas, and an upper bound $\fsp \lesssim f_a=10^3$ TeV we find
\begin{equation}
    m_{\eta} \leq 10 \, \text{GeV}.
    \label{eq:eta_mass_bound}
\end{equation}
As mentioned in Section \ref{sec:global_symmetries}, the $\eta_B$ remains perfectly massless, as it is associated with a $U(1)_B$ baryon-number symmetry which is not explicitly broken by the Yukawa interaction. In this case, as pointed out in \cite{Valenti:2022tsc}, $\eta_B$ will behave like a photophobic axion-like particle. In order to avoid phenomenological contraints, one solution would be to gauge $U(1)_{B-L}$ to gauge it away.

\subsection{Sources of misalignment and of CP violation}
\label{sec:cp_violation}
Given the discussion so far, our solution to the strong CP problem could still be spoiled if the two topological angles in \eqref{eq:lag_topological_term} would be misaligned by more than a part in $10^{10}$. If this happens, a single axion would not be able to cancel both terms and actually, being the confinement scale of $\Sp$ higher that the one of $\Suc$, the axion would actually tend to align with the topological angle of the first sector, thus not solving the QCD strong CP problem. 
As discussed in the introduction, the tree-level relation $\theta_{\mathrm{c}}=\theta_{\mathrm{Sp}}$ is guaranteed by the fact that $\Suc$ and $\Sp$ are embedded in \textit{grandcolor}. The task is now to ensure that no radiative effects spoil it. Assuming that no tree-level coupling between the new scalars $\Phi,\Xi$ and the SM-like ($Q, \,U, \,D$) fermions exist\footnote{These interactions can be forbidden via the underlying gauge structure of the scalar composite dynamics, or by charging them under additional gauge groups.}, the radiative misalignment from the Yukawa sector is completely negligible due to the high loop suppression \cite{Ellis:1978hq, Jarlskog:1985ht}. Contributions from the scalar potential $V(\Phi, \Xi)$ can also be avoided by choosing the parameters to not trigger spontaneous CP breaking. As pointed out in \cite{Valenti:2022tsc}, beyond these contributions, the first non-renormalizable operators that could lead to a misaligned of the two topological terms are 
\begin{equation}
    \frac{c_5}{f_{\text{ETC}}}\frac{g_{\text{GC}}^2}{32\pi^2}\Tr \left[\Phi G_{\text{GC}}\tilde{G}_{\text{GC}} \right]+  \frac{c_6}{f_{\text{ETC}}^2}\frac{g_{\text{GC}}^2}{32\pi^2} \Tr \left[ \Phi^{\dagger}\Phi G_{\text{GC}}\tilde{G}_{\text{GC}} \right] + \frac{c_6 ^ \prime}{f_{\text{ETC}}^2}\frac{g_{\text{GC}}^2}{32\pi^2} \Tr \left[\Xi^{\dagger}\Xi G_{\text{GC}}\tilde{G}_{\text{GC}} \right],
\end{equation}
where the trace is intended over all possible index contractions consistent with gauge invariance and we have identified $f_{\text{ETC}}$ as the UV cut-off of our theory, which must lie below the Planck scale. The first of these terms implies  $\theta_{\mathrm{c}}-\theta_{\mathrm{Sp}} \sim f_{\text{GC}}/f_{\text{ETC}}$, so that to respect \eqref{eq:theta_relation} up to corrections of order $10^{-10}$, we have the constraint $f_{\text{GC}} \lesssim 10^{-10} f_{\text{ETC}}$. From the two dimension 6 operators we get instead $f_{\text{GC}}\lesssim 10^{-5} f_{\text{ETC}}$. The first bound can be avoided if we charge $\Phi$ under an additional gauge group while the second one could also be removed by choosing the right underlying structure of the composite scalars. We also notice that both operators are further suppressed by the grandcolor gauge coupling, which at the scale $f_{\text{GC}}$ is expected to be small.

\section{Phenomenology}
\label{sec:phenomenology}
The considerable number of pNGBs in our model makes its phenomenology very rich but also complex to analyze. The scope of this section is to investigate the {\it key phenomenological signatures}, focusing in particular on the axion. The phenomenology of the composite Higgs sector has been broadly analyzed in the literature, see e.g. \cite{Cacciapaglia:2020kgq,Cacciapaglia:2022zwt} for reviews, the one for the $\pi_{\psi}$ in \cite{Valenti:2022tsc}, where the rough bound $\fsp>1$ TeV was found, while the phenomenology of the $\pi_{\chi}$ as well as $\sigma_{\omega}, \sigma_{\chi}$ has been studied in  detail in~\cite{Belyaev:2016ftv, Cacciapaglia:2021agf, Cacciapaglia:2015eqa}. Following the bounds of these works, we simply adopt a conservative assumption on the masses of the resonances of $M_{\pi, \sigma} > 1.5$\,TeV. As discussed in Section \ref{sec:chiral_lagrangian}, these are expected to have a mass or order $f_a$.

\begin{figure}[t]
    \centering
    \includegraphics[scale=1.3]{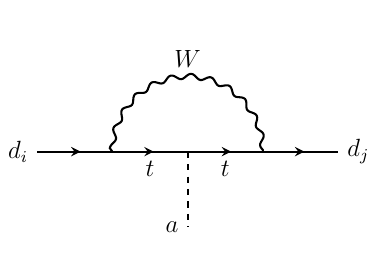}
    \caption{FCNC mediated by the singlet. Figure from \cite{Esser:2023fdo}. }
    \label{figure_fcnc}
\end{figure}

Depending on the mass of $\eta$, most relevant experimental constraints come either from collider searches \cite{Blasi:2023hvb, Bauer:2017ris} or from flavor physics \cite{Freytsis:2009ct, Izaguirre:2016dfi, Bjorkeroth:2018dzu,Ziegler:2019gjr,Bauer:2021mvw, Esser:2023fdo}. In our case, given the rough upper bound on the singlet mass presented in \eqref{eq:eta_mass_bound}, we will focus on the latter. The presence of a light state coupled to SM fermions can be investigated via the measurement of rare meson decays, in particular the one of Kaons and B-mesons. For the axion, the relevant decay channels are
\begin{equation}
    K \rightarrow \pi + a, \qquad B \rightarrow K + a,
    \label{KB_decay}
\end{equation}
which are generated by the diagram in Figure \ref{figure_fcnc}. The best limits come from BaBar  searches of B-meson decays into Kaons and invisibles, which find, at $90\%$ C.L.~\cite{BaBar:2013npw} 
\begin{equation}
    \text{BR}(B\rightarrow K + \text{inv}) < 3.2 \times 10^{-5}  \quad \text{for $m_{\eta}\leq 5$ GeV}.
\end{equation}
This upper bound on the branching ratio can then be translated into an upper bound on the axion-top coupling \cite{Esser:2023fdo, Izaguirre:2016dfi,Ball:2004ye} 
\begin{equation}
    \abs{\frac{c_t}{f_a}} \leq 1.5\times10^{-6} \, \text{GeV}^{-1} \text{  for $m_{\eta}\leq 5$ GeV  }.
\end{equation}
We now analyze two scenarios. In the first, the two lightest $\psi$ fermions are assumed to have masses comparable to the ones of the SM up and down quarks, whereas in the the second, they are assumed to have a mass similar to those of the second generation of SM quarks. Similarly to the case for the exotic partner of the top, $\psi_t$, for which we had to assume a smaller Yukawa coupling, this can be obtained if the Yukawas of the SM quarks and of their exotic partners are different. We use now the conservative bounds
\begin{align}
    \fsp & \gtrsim 1 \text{ TeV} , \\
    f_a & \gtrsim 2 \text{ TeV} , \\
    \fsp & \lesssim f_a ,
\end{align}
where the first one comes from the decays of $\pi_{\psi}$, the lower bound on $f_a$ is a conservative one to avoid experimental constraints on the composite nature of the Higgs and the third one reflects a consistency condition dictated by the way we have analyzed the potential in the previous sections. The inverse relation is in principle possible, but would require a different treatment of the potential and a different analysis compared to the one we have conducted here.
With the bounds we have just discusses, the axion mass obtained in \eqref{eq:eta_mass_expression} is then constrained within the band
\begin{equation}
    y_{\psi_1} y_{\psi_2} \frac{(1 \text{TeV})^4}{f_a^2 } \lesssim m_{\eta}^2 \lesssim y_{\psi_1} y_{\psi_2} f_a^2.
    \label{axion_bounds_eq}
\end{equation}
In the expression above, $y_{\psi_{1,2}}$ indicates the Yukawa couplings of the two lightest $\psi$ fermions, in complete analogy with the expression for the canonical QCD axion, where its mass depends on the up and down mass.
\begin{figure}[t]
    \centering
    \includegraphics[scale=0.37]{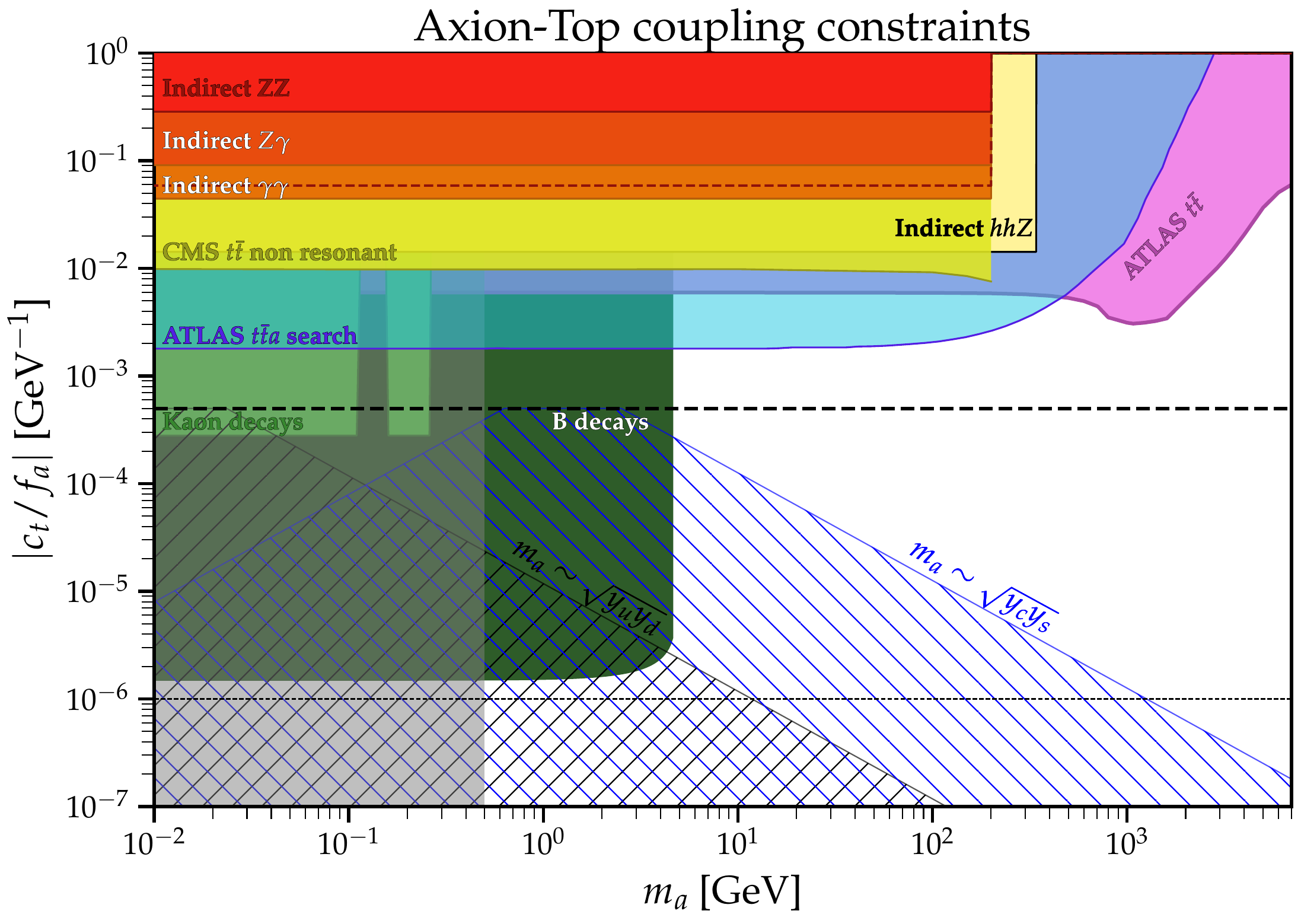}
    \caption{Main bounds on the axion-top couplings. The black dashed area indicates the parameter space of our model when we assume that the first-generation $\psi$ fermions have the same Yukawas as the up and down quarks. The blue dashed area assumes instead Yukawas similar to the second generation.  The thick black dashed line indicates $f_a=2\,$TeV, while the thin one indicates $f_a=10^3\,$TeV, motivated by tuning arguments. The gray shaded region is conservatively neglected due to possible bounds on the induced axion-photon coupling. Adapted from \cite{Esser:2024pnc}.}
    \label{figure_exclusion_plot}
\end{figure}
In Figure \ref{figure_exclusion_plot} we have reported the current bounds on the axion-top couplings and the region of parameter space of interest for our model. As we can see, the parameter space is already heavily constrained in the case of Yukawa couplings similar to the first generation ones. In this case, the only region that seems to be viable is the one corresponding to $f_a\sim 10^3$ TeV, but such a high scale for the composite sector would reintroduce a small hierarchy problem. 

The second benchmark scenario of somewhat larger $\psi$ Yukawa couplings is instead more promising. An axion with a mass of about $10$ GeV can be achieved without excessive tuning in the Higgs sector. Nevertheless, this scenario also features some challenges. We have seen that to avoid a potentially large mixing between the Higgs-like pions $\pi_{\psi}^H$ and the bidoublet $H$, the heaviest exotic quark must be lighter than the bottom. If we further assume that the two lightest ones have a Yukawa comparable to the charm and strange, we conclude that the mass spectrum of the $\psi$ fermions must be less hierarchical than the one of the SM quarks. This in turn will modify the minimization of the axion potential and in particular could introduce a larger mixing between $\eta$ and the neutral pions $\pi_{\psi}^0$, thus modifying its coupling to the top and to photons. We notice here that non-hierarchical Yukawas for the $\psi$ sector are in principle not a problem, they would just require a more detailed analysis of the $\eta$, its mixing with the other pNGBs and its couplings. 

With this in mind, a more realistic scenario might be an intermediate one between the two just discussed. In particular, if the two lightest $\psi$ quarks had a mass higher than that of the up and down but were still lighter than the second generation, all these problems could be avoided and an axion with a mass of around 10 GeV would still be achievable with a moderate tuning in the Higgs sector. Future collider searches and flavor experiments have the potential to discover such an axion related to Higgs naturalness.

\section{Conclusions}
We presented a model that solves the strong CP problem and the hierarchy problem simultaneously. Our setup features a composite axion arising from the same minimal coset that realizes a composite Higgs from fundamental gauge-fermion dynamics, the latter solving the hierarchy problem without introducing excessive tuning. Since such a solution requires the common Higgs-axion decay constant to reside not much above the TeV scale, which would be strongly excluded for a standard QCD-like axion, we included further contributions to the axion mass via additional small instanton contributions from an enlarged color group.
We demonstrated that the CP-odd scalar singlet contained in the $SU(4)/Sp(4)$ coset of the fundamental composite Higgs can cancel the $\theta$ angle of both $SU(3)_c$ and $Sp(2\Ngc)$ of the enlarged color sector, thereby dynamically solving the strong CP problem, while maintaining a low compositeness scale for the Higgs boson. Beyond showing that the axion potential can be protected from contributions that would misalign the minimum, we confirmed that the relation $\theta_{c}=\theta_{\mathrm{Sp}}$ is stable and that sources of problematic CP violation can be controlled.
Finally, the viable parameter space that we identified illustrates that future collider searches and precision measurements of meson decays have the potential to discover this modified QCD axion, emerging from Higgs naturalness.

\acknowledgments

We are grateful to Andreas Bally, Anke Biek{\"o}tter, Giacomo Cacciapaglia, Yi Chung, Gabriele Ferretti, Maya Hager, and Francesco Sannino for useful discussions.

\newpage
\appendix
\section{Group theory factors}

\begin{table}[h]
    \centering
    \renewcommand{\arraystretch}{1.2} 
    \setlength{\tabcolsep}{6pt} 
    \centering
    \begin{tabular}{|c|c|c|c|}
        \hline
        R     & T(R)  & $C_2(R)$ & d(R) \\ \hline \hline
        $\mathbf{F}$    & $1/2$ & $(N^2-1)/2N$ & $N$\\
        $\mathbf{Adj}$     & $N$ &  $N$& $N^2-1$ \\ \hline
    \end{tabular}
    \captionsetup{justification=centering}
    \caption{Group theory factors for $SU(N)$.}
    \label{tab:dynkin_casimir_1}
    \vspace{5mm}
    \centering
    \begin{tabular}{|c|c|c|c|}
        \hline
        R     & T(R)  & $C_2(R)$ & d(R) \\ \hline \hline
        $\mathbf{F}$    & $1/2$ & $(2N+1)/4$ & $2N$\\
        $\mathbf{Adj}$     & $N+1$ & $N+1$& $N(2N+1)$\\
        $\mathbf{A}_2$     & $N-1$ & $N$ & $N(2N-1)-1$ \\ \hline
    \end{tabular}
    \captionsetup{justification=centering}
    \caption{Group theory factors for $Sp(2N)$.}
    \label{tab:dynkin_casimir_2}
    \vspace{5mm}
    \centering
    \begin{tabular}{|c|c|c|c|}
        \hline
        R     & T(R)  & $C_2(R)$ & d(R) \\ \hline \hline
        $\mathbf{F}$    & $1$ & $(N-1)/2$ & $N$\\
        $\mathbf{Adj}$     & $N-2$ &  $N-2$ & $N(N-1)/2$ \\
        $\mathbf{Spin}$ (even)  & $2^{\frac{N-8}{2}}$ & $N(N-1)/16$ & $2^{\frac{N-2}{2}}$ \\
        $\mathbf{Spin}$ (odd)  & $2^{\frac{N-7}{2}}$ & $N(N-1)/16$ & $2^{\frac{N-1}{2}}$ \\ \hline
    \end{tabular}
    \captionsetup{justification=centering}
    \caption{Group theory factors for $SO(N)$.}
    \label{tab:dynkin_casimir_3}
\end{table}

\bibliographystyle{JHEP}
\bibliography{biblio.bib}

\end{document}